\documentclass[report,preprint,a4paper]{revtex4}%
\usepackage{graphics}
\usepackage{graphicx}
\usepackage{bm}
\usepackage{amsmath}
\usepackage{mathtools}
\usepackage{color}
\usepackage{bbold}
\usepackage{braket}
\usepackage{MnSymbol}
\usepackage{upgreek}
\usepackage{leftidx}

\begin{document}
\title{Local Photons}
\author{Daniel Hodgson, Jake Southall, Rob Purdy and Almut Beige}
\affiliation{The School of Physics and Astronomy, University of Leeds, Leeds LS2 9JT, United Kingdom.}

\begin{abstract}
The classical free-space solutions of Maxwell's equations for light propagation in one dimension include wave packets of any shape that travel at the speed of light. This includes highly-localised wave packets that remain localised at all times. Motivated by this observation, this paper builds on recent work by Southall {\em et al.} [J.~Mod.~Opt.~{\bf 68}, 647 (2021)] and shows that a local description of the quantised electromagnetic field, which supports such solutions and which must overcome several no-go theorems, is indeed possible. Starting from the assumption that the basic building blocks of photonic wave packets are so-called {\em bosons localised in position} (blips), we identify the relevant Schr\"odinger equation and construct Lorentz-covariant electric and magnetic field observables. In addition we show that our approach simplifies to the standard description of quantum electrodynamics when restricted to a subspace of states.
\end{abstract}

\maketitle

\section{Introduction} \label{sec1}

The problem of describing single-photon states in the position representation has been a long-standing challenge to physicists.  As early as 1948, Pryce \cite{Pryce} discussed how, in the case of a photon, there seems to be no possibility of defining a three-dimensional position operator with commuting components.  An equivalent statement is, there can be no wave function for the photon which is localised in all three dimensions at once.  Only a year later, another by now well known paper by Newton and Wigner \cite{Wigner} proved that such a position operator cannot exist for massless particles with a spin greater than one half (the photon is a spin-1 particle) if the eigenstates of the position operator, or localised particles, are assumed to have a spherical symmetry.  More recently, Hawton and Debierre \cite{Hawton7} noticed that it would be more accurate to say that the photon position operator must have a cylindrical symmetry rather than a spherical one due to the divergence condition on free electromagnetic (EM) fields.

The continued research into photon position operators has provided more detailed analyses and simpler proofs of the localisation problem. Examples are the proof provided by Jordan \cite{Jordan} and the investigations of Fleming \cite{Fleming1,Fleming3,Fleming2} and Halvorsen \cite{Halvorsen2}. Possible alternative conditions for a position operator have also been studied (see e.g.~Ref.~\cite{Shojai}). Several distinct approaches to building a position-dependent description of the photon have also come about in this time. In one instance, Wightman \cite{Wightman} reformulated the work of Newton and Wigner in the framework of imprimitive representations of the Euclidean group.  Wightman similarly concluded that a photon could not be localised. Later, Jauch and Piron \cite{JandP} generalised some of the axioms in Wightman's scheme, developing the notion of weak localisability.  Amrein \cite{Amrein} showed that combinations of photons of different helicity are weakly localisable. Other authors aimed at constructing spin-1 divergence-less single-photon wave functions whose squared moduli represent some useful and measurable physical quantity. 

For example, Hawton \cite{Hawton1} made progress in this direction by demonstrating that it is possible to construct transversely polarised localised photon wave packets provided that one also takes into consideration the longitudinally polarised components of the momentum wave function (see also Refs.~\cite{Hawton2}).  The position operator obtained in this way differed from Pryce's earlier position operator by a Berry connection term.  Amongst others, Bia\l ynicki-Birula \cite{BB2,BB3}, Sipe \cite{Sipe} and Raymer and Smith \cite{Raymer} constructed both first and second quantised solutions of the massless Dirac equation and obtained wave functions that are locally related to the Riemann-Silberstein vector and, therefore, the electric and magnetic field observables. It is often believed that a local relationship to the field observables is a necessary condition for any physically significant wave function; perhaps this view was instigated by the form of the Glauber photo-detection operators \cite{Glauber1}.  In the view of Knight \cite{Knight} and Licht \cite{Licht}, a state can only be localised if a measurement of either the electric or magnetic field at some other location view the system as in its ground state. From this point of view, when the field observables do not commute, single photon states cannot be localised \cite{BB1}.

When a prospective wave function is locally related to the field observables, the typical square Born rule now provides an energy rather than a probability causing further difficulties for the interpretation of the wave function. There are two methods of circumventing this problem. One method is to introduce a modified inner product that has the correct dimensions.  This can be done either by normalising the photon wave function with respect to photon energy, as is done in Refs.~\cite{BB3} and \cite{Gross}, or by treating the system as a biorthogonal system \cite{Raymer,Glauber1,Hawton3,Hawton4, Brody,Hawton5,Jaromir}. For further reading on biorthogonal systems see, for example, Refs.~\cite{Mostafazadeh2, Mostafazadeh1}. This approach introduces a non-standard inner product that normalises the wave functions by a term with units of energy.  The inner product between field states then has the typical units of probability density, and may therefore retain its usual probabilistic interpretation.  A second and simpler alternative is to consider excitations of the correct units as physical, regardless of their relation to the field observables. This approach was adopted in the development of the Landau-Peierls wave function \cite{Landau and Peierls} which has been criticised for its non-local transformation properties; however, it has since been revived by Cook \cite{Cook1,Cook2} and Mandel \cite{Mandel} who have constructed second quantised, position-dependent excitations.

In spite of arguments against such excitations, in this paper we shall follow a similar second quantised approach to photon localisation that avoids biorthogonal quantum physics. We shall assume that  the states of photons localised at different locations are mutually orthogonal to one another such that a photon localised at a position $x$ cannot be found at $x' \neq x$ and vice versa. In this way, we obtain a theory in which the likelihood of a photon being found in a certain region of space can be calculated by means of a projection operator. Moreover, as we shall see below, this assumption implies that the annihilation and creation operators of localised photons have bosonic commutator relations. These are in good agreement with linear optics experiments with ultra-broadband photons, which confirm the bosonic nature of these localised photonic particles \cite{exp,exp2,exp3,exp4}.

As we shall see below, our scheme has many similarities with previous work by Bennett {\em et al.}~\cite{Bennett} which takes a shortcut to the introduction of particle annihilation and creation operators. Instead of first verifying their possible existence by establishing a harmonic oscillator Hamiltonian, the existence of photonic particles that are the basic building blocks of travelling waves is simply postulated and the properties of the corresponding fields are derived by demanding consistency with classical electrodynamics. The main difference of the approach that we present here is that we treat the local solutions of Maxwell's equations, rather than the monochromatic solutions, as the basic building blocks of the EM field. Our approach also has some similarities with the approach by Ornigotti {\em et al.} \cite{Szameit} which quantises so-called X waves \cite{book} instead of monochromatic waves which are  diffraction- and dispersion-free solutions of Maxwell's equations. Moreover, Aiello \cite{Aiello,Aiello2} recently obtained a phenomenological, non-standard description of the EM field by considering the monochromatic solutions of the Helmholtz wave equations and a paraxial wave equation for light propagation in free space.

However, the localisation of single photons results in another problem \cite{Halvorsen1,Hegerfeldt2,Hegerfeldt3,Skagerstam,Perez,Malament}. In a paper published in 1974, Hegerfeldt \cite{Hegerfeldt2,Hegerfeldt3} provided a short proof that, if the probability of detecting a particle in a certain region of space is given by the expectation value of some suitably chosen projection operators, then that same particle will spread out superluminally. A similar proof was also found by Malament \cite{Malament}. The only assumptions made in the derivation of this superluminal spreading is that the particle Hamiltonian is translation invariant and bounded from below. More recently, it has been shown that the sole cause of the spreading is the lower bound placed on the Hamiltonian of the system \cite{Hegerfeldt1,Hegerfeldt4}.  

The problem of superluminal spreading described above also lies at the heart of a problem which emerged after Fermi calculated, in 1932 \cite{Fermi}, the minimum time for a ground-state atom to transition into an excited state through the absorption of radiation emitted by a second nearby atom. As one would intuitively expect from causality considerations, he found that there is a zero probability for this transition to happen until enough time has elapsed for light to propagate from one atom to the other. Much later, however, Shirokov \cite{Shirokov} pointed out that this causal result relied upon an approximation made in the original derivation.  The particular nature of the apparent non-causal contributions generated in Fermi's problem have since been investigated and discussed in a number of different contexts \cite{Rubin,Biswas,Milonni1,Borrelli}. In 1994, Hegerfeldt demonstrated how, under quite general assumptions, the second atom may be excited after arbitrarily short times \cite{Hegerfeldt1}. This conclusion was repudiated by Buchholz and Yngvason \cite{Yngvason} who argued that, due to the hyperbolicity of the relevant equations of motion, a measurement at the second atom cannot learn anything about the first atom until a sufficient amount of time has elapsed such that causality is preserved. Moreover, using a magnus expansion of the time-evolution operator, Ben-Benjamin \cite{Ben-Benjamin} has shown that causality is maintained in the strictest sense. 

\begin{figure}[t]
\begin{center}
\includegraphics[width = 0.8\textwidth]{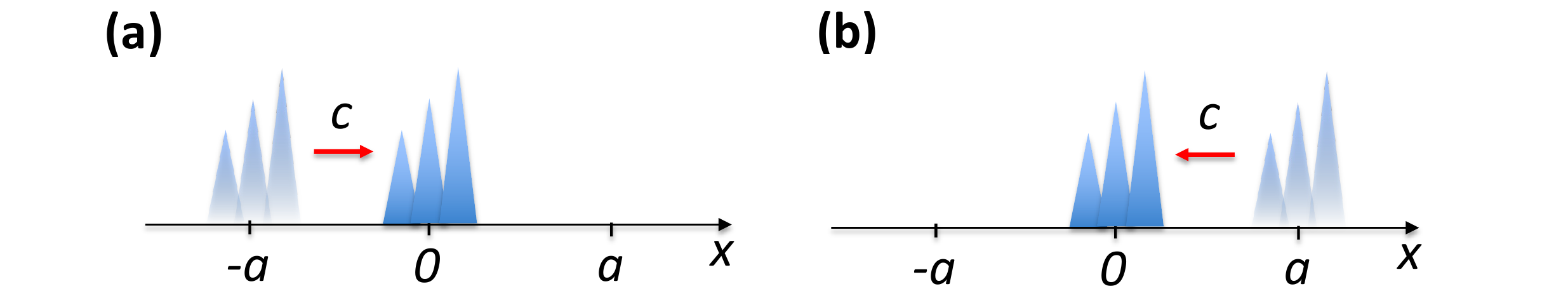}
\end{center}
\caption{The figures illustrate two different scenarios in which a single-photon wave packet of a certain shape travels  at the speed of light $c$ in a well-defined direction. (a) Here the wave packet is initially placed in the vicinity of the point $x=-a$ and moves to the right. (b) Here the wave packet is located at $x=a$ and moves to the left. At $t=0$, the wave packets in both scenarios are easily distinguishable and hence correspond to pairwise orthogonal states (cf.~Eq.~(\ref{2})). However  after some time $t=a/c$, the wave packets reach the point $x=0$ in both scenarios and their wave functions seem to be the same in both cases (cf.~Eq.~(\ref{2})), up to an overall phase factor. From this we conclude that the initial states propagate unitarily only if their respective state vectors belong to separate Hilbert spaces, which we label by a parameter $s$ identifying their direction of propagation. In the following, $s=-1$ and $s=1$ correspond to left- and to right-moving wave packets respectively.} \label{fig:1}
\end{figure}

Notwithstanding the conclusions reached by the above investigations, the superluminal spreading described by Hegerfeldt and Malament is a significant obstacle to the introduction of local single-photon wave functions $\psi(x,t)$. The problem can be traced back to the fact that, in current theories, localised photons are incapable of displaying characteristics that are unmistakably present in the solutions of Maxwell's equations in free space. From classical electrodynamics we know that it is possible to generate wave packets of light of any shape which propagate at the speed of light, i.e.~without dispersion \cite{book}. To overcome this issue, let us simply assume for a moment that single-photon wave functions $\psi(x,t)$ exist and determine their respective properties. In this way, we can identify which alterations have to be made to the current standard descriptions of light \cite{Bennett} in order to support the existence of local photons.

In the following, we consider the two different scenarios illustrated in Fig.~\ref{fig:1}. In the first scenario, a right-moving single-photon wave packet with an initial wave function $\psi_1(x,0)$ is placed near the point $x=-a$ (cf.~Fig.~\ref{fig:1}(a)). In the second scenario, a left-moving single-photon wave packet with an initial wave function $\psi_2(x,0)$ is placed near $x=a$  (cf.~Fig.~\ref{fig:1}(b)). If both wave packets have the same shape, $\psi_1(x,0)$ and $\psi_2(x,0)$ differ at most such that 
\begin{equation}
\psi_1(x,0) = \psi_2(x+2a,0) \, {\rm e}^{{\rm i} \varphi} 
\end{equation}
where $\varphi$ denotes a phase. Otherwise, the probability density of finding the first photon at $x$ and the probability density of finding the second photon at $x+2a$ would not be the same. If the two wave packets do not overlap in space, they are easily distinguishable and their state vectors must be pairwise orthogonal, 
\begin{equation} \label{2}
\braket{\psi_1(0)|\psi_2(0)} = 0 \, .  
\end{equation}
From this it follows that the overlap between the state vectors remains zero at all times and  
\begin{equation} \label{3}
\braket{\psi_1(t)|\psi_2(t)} = \braket{\psi_1(0)|U^\dagger(t,0) U(t,0)|\psi_2(0)} = \braket{\psi_1(0)|\psi_2(0)} = 0\,  , 
\end{equation}
if both state vectors evolve unitarily with a time-evolution operator $U(t,0)$ with $U^\dagger(t,0) U(t,0) = 1$. However, this is not the case: at the time $t = a/c$ the wave packets overlap and can no longer be distinguished.  Their wave functions differ at most by a phase factor since 
\begin{equation} \label{4}
\psi_1(x,t) = \psi_1(x-ct,0) = \psi_2(x-ct+2a,0) \, {\rm e}^{{\rm i} \varphi} = \psi_2(x,t) \, {\rm e}^{{\rm i} \varphi}\,  .
\end{equation}
For Eq.~(\ref{3}) to hold at all times, the quantum states of left- and right-moving wave packets must belong to separate Hilbert spaces. This means they must be characterised by an additional degree of freedom, namely their direction of propagation. In the following, we therefore proceed as originally proposed by Dirac \cite{Dirac} and also previously discussed in Ref.~\cite{Jake}. More concretely, we distinguish four types of photons and add indices $s \lambda$ to their wave functions with $s=-1$ and $s=1$ indicating left- and right-moving photons respectively while $\lambda = {\sf H}, {\sf V}$ denotes their polarisation.

Next we notice that all photons with a well-defined direction of propagation $s$ and polarisation $\lambda$ travel at the speed of light, $c$ such that
\begin{equation}
\psi_{s \lambda}(x,t) = \psi_{s \lambda}(x-sct,0)\,  .
\end{equation}
Replacing the wave functions in the above equation by superpositions of their respective momentum space wave functions $\widetilde{\psi}_{s \lambda}(k,t)$, as suggested by the Fourier transform, one can now show that
\begin{equation} \label{6}
\int_{-\infty}^\infty \text{d}k \, \widetilde{\psi}_{s \lambda}(k,t) \, {\rm e}^{\pm{\rm i} kx}  
=  \int_{-\infty}^\infty \text{d}k \, \widetilde{\psi}_{s \lambda}(k,0) \, {\rm e}^{\pm{\rm i} k(x-sct)} \,  .
\end{equation}
The fact that this relation must hold for wave packets of any shape implies that
\begin{equation}
\widetilde{\psi}_{s \lambda}(k,t)  =  \widetilde{\psi}_{s \lambda}(k,0) \, {\rm e}^{\mp{\rm i} sckt} \, ,
\end{equation}
where the sign in the exponent depends on which Fourier transform has been used in Eq.~(\ref{6}). Such dynamics can be generated by the Schr\"odinger equation of a collection of harmonic oscillators but, as previously observed in Ref.~\cite{Jake}, the corresponding Hamiltonian must have positive and negative eigenvalues. Notice also that for any wave packet with only positive or only negative wave numbers in its spectrum, the sign of the Fourier transform can always be chosen such that only a Hamiltonian with positive energy eigenvalues is required. However, this no longer applies for wave packets with positive and negative $k$ and a well-defined direction of propagation $s$.

The arguments presented above illustrate how states characterised only by position and polarisation, as they are in the current theory of the quantised EM field (cf.~e.g.~Ref.~\cite{Bennett}), cannot describe the unitary evolution of localised wave packets. In addition, they show that a complete description of the quantised EM field requires a system Hamiltonian which no longer coincides with the energy observable of the EM field which only has positive eigenvalues. In this paper these observations are taken into account in an alternative approach to quantising the free EM field for light propagation in one dimension. Our starting point will be the initial assumption that basic building blocks of light are single photons that can be localised without becoming necessarily dispersive. As in Ref.~\cite{Jake}, we refer to these localised photons in the following as {\em bosons localised in position} (blips). 

There are five sections to this paper. In Section \ref{Sec:position}, we shall  introduce the equation of motion in position space and construct a Hilbert space containing a complete set of mutually orthogonal position states that evolve without dispersion. We shall then define the usual set of EM field observables up to an overall re-scaling operator $\mathcal{R}$ that obey Maxwell's equations and act upon this Hilbert space.  We shall further define a Hamiltonian operator that generates the unitary dynamics of these states. In Section \ref{Sec:momentum}, we shall again construct a Hilbert space for the free EM field, now in the momentum representation. A corresponding set of field observables will also be constructed. Afterwards, in Section \ref{sec:relationship}, we establish a connection between the two representations, which share the vacuum state. It is shown that the position- and the momentum-space annihilation operators can be linked via a Fourier transform in much the same way as we link local and non-local electric and magnetic field amplitudes in classical electrodynamics. Moreover, we identify the re-scaling operator $\mathcal{R}$ by imposing Lorentz covariance. In this way, the momentum representation of this more complete theory may be compared with the traditional theory of the quantised EM field, which is typically expressed in the momentum representation. Finally, we summarise our findings in Section \ref{Sec:photonsdiscussion}. 

\section{Quantisation in the position representation}
\label{Sec:position}

\subsection{Classical Electromagnetism}

\label{Sec:classical1}

The theory of electromagnetism in one dimension is concerned with the properties and dynamics of two fundamental quantities: the electric field ${\boldsymbol E}(x,t)$ and the magnetic field ${\boldsymbol B}(x,t)$. These two real fields are vector valued, having components in all three space dimensions, and are parametrised by a position along a single axis $x$ and a time $t$, which represent the position and time at which the fields are measured. The dynamics of the electric and magnetic fields are governed by Maxwell equations. In a system in which there are neither free charges nor free currents, which we call free space, Maxwell's equation are given by
\begin{eqnarray}	\label{fMaxwell1}
&&	\mathbf{\nabla}\cdot {\boldsymbol E}(x,t) = 0 \, , ~~
\mathbf{\nabla}\times{\boldsymbol E}(x,t) = -\frac{\partial}{\partial t}{\boldsymbol B}(x,t) \, , \notag \\
&&	\mathbf{\nabla}\cdot {\boldsymbol B}(x,t) = 0 \, ,  ~~
\mathbf{\nabla}\times{\boldsymbol B}(x,t) = \frac{1}{c^2}\frac{\partial}{\partial t}{\boldsymbol E}(x,t) \, . 
\end{eqnarray}
These fields are known as the free fields.

The above equations are not independent, but couple together different components of both the electric and magnetic field vectors.  By manipulating Maxwell's equations we can determine the six independent second-order equations
\begin{eqnarray}
	\label{1DMaxwell1}
	\left( \frac{\partial^2}{\partial x^2} - \frac{1}{c^2}\frac{\partial^2}{\partial t^2} \right) O_i(x,t) 
	= \left(\frac{\partial}{\partial x} - \frac{1}{c}\frac{\partial}{\partial t}\right)\left(\frac{\partial}{\partial x} + \frac{1}{c}\frac{\partial}{\partial t}\right)O_i\mathbf(x,t) 
	= 0 \, .
\end{eqnarray}
Here $O_i(x,t)$ represents any of the six components of the ${\boldsymbol E}(x,t)$ or ${\boldsymbol B}(x,t)$ vectors in a Cartesian basis.  The constant $c$ is the speed of light.  By imposing the divergence conditions of Eq.~(\ref{fMaxwell1}) on the one-dimensional fields, one finds that the $x$ components of the electric and magnetic fields are constant and, by choice, vanishing.  For a free field propagating in the $x$ direction only, the electric and magnetic field vectors have components in the $y$ and $z$ directions only.

In Eq.~(\ref{1DMaxwell1}), the one-dimensional wave equation can be factorised into a product of two, first-order differential operators. The most general solution of this equation, therefore, is a superposition of the field vectors that vanish when acted upon by the differential operators in either set of brackets.  After taking into account the allowed polarisations of the one-dimensional solutions, one finds that the general solution for the electric field vector is of the form
\begin{eqnarray}
	\label{Efield1}
	{\boldsymbol E}(x,t) = \sum_{s = \pm 1} E_{s{\sf H}}(x,t) \, \hat{\textbf{y}} + E_{s{\sf {\sf V}}}(x,t) \, \hat{\textbf{z}}
\end{eqnarray}
where $E_{s\lambda}(x,t)$ with $s = \pm1$ satisfies the first-order differential equation
\begin{eqnarray} \label{11}
	\left(\frac{\partial}{\partial x} + \frac{s}{c}\frac{\partial}{\partial t}\right) {\boldsymbol E}_{s\lambda}(x,t) = 0 \, .
\end{eqnarray}
Moreover, $\hat{\textbf{y}}$ and $\hat{\textbf{z}}$ are unit vectors that lie parallel to the $y$ and $z$ axes respectively and are oriented in the direction of the increasing coordinate. The solutions of this first-order equation have a dependence on the space-time distance $x-sct$ only and $E_{s\lambda}(x,t) = E_{s\lambda}(x-sct)$.  Using either of the curl conditions in Eq.~(\ref{fMaxwell1}), one may show that the magnetic field is given by
\begin{eqnarray}
	\label{Bfield1}
	{\boldsymbol B}(x,t) = \sum_{s = \pm 1} \frac{s}{c} \, \left( - E_{s{\sf V}}(x,t) \, \hat{\textbf{y}} + E_{s{\sf H}}(x,t) \, \hat{\textbf{z}}\right) \, .
\end{eqnarray}
The electric and magnetic field vectors ${\boldsymbol E}(x,t)$ and ${\boldsymbol B}(x,t)$ are not independent but are mutually propagating and orthogonal to each other.  To solve ${\boldsymbol E}_{s\lambda}(x,t)$ exactly, one will need to impose a set of boundary conditions on the fields and their time derivatives at some chosen time. In one dimension, the energy of the EM field is given by the expression
\begin{eqnarray}
	\label{classicalenergy1}
	H_{\text{energy}}(t) = \int_{-\infty}^\infty\text{d}x \, \frac{\varepsilon_0A}{2} \, \left( {\boldsymbol E}(x,t)^2 + c^2 \, {\boldsymbol B}(x,t)^2\right) \, .
\end{eqnarray}
Here $\varepsilon_0$ is the electric permittivity of free space and $A$ denotes the area that the fields occupy in the $y$-$z$ plane. The total energy of the system is a constant of motion.

\subsection{Quantisation in position space}
\label{Sec:1Dposition}

\subsubsection{Blip states}
\label{Sec:blipstates}

In the past we have assumed that the basic building blocks of the EM field are monochromatic photons. Let us now take a different approach and assume that the fundamental excitations are a set of spatially localised photonic wave packets that propagate along the $x$ axis of a Cartesian coordinate system.  We shall call these localised excitations blips, which stands for {\em bosons localised in position}.  At any given time $t$, a blip state can be fully characterised by its position $x$ along the $x$ axis, a polarisation $\lambda$ and a direction of propagation $s$. As mentioned in the Introduction, $s$ takes values $\pm 1$ with $s = +1$ indicating propagation in the direction of increasing $x$ and $s = -1$ indicating propagation in the direction of decreasing $x$.  As is usual in one dimension, $\lambda = {\sf H}, {\sf V}$.

In the following, we denote the annihilation operators for these blip excitations $a_{s\lambda} (x,t)$ in the Heisenberg picture and $a_{s\lambda} (x)$ in the Schr\"odinger picture. To identify a Hilbert space, we proceed as usual and first introduce a vacuum state $\ket{0}$ for the EM field. The vacuum state is annihilated by the annihilation operators $a_{s\lambda}(x,t)$ for all $x$, $t$, $s$ and $\lambda$: 
\begin{eqnarray} 
	\label{annihilate1}
	a_{s\lambda}(x,t) \, \ket{0} = 0 
\end{eqnarray}
and should be normalised such that $\braket{0|0} = 1$. As we shall see below, it is also the minimum energy state of the EM field. The Hermitian conjugate of $a_{s\lambda}(x,t)$, i.e.~$a^\dagger_{s\lambda}(x,t)$, is the creation operator that generates a single-blip excitation state when applied to $\ket{0}$:
\begin{eqnarray}
	\label{blipstate1}
	\ket{1_{s\lambda}(x,t)} = a^\dagger_{s\lambda}(x,t)\ket{0} \, .
\end{eqnarray}
By repeatedly applying blip creation operators to the vacuum, we are able to generate a complete set of multi-particle states that eventually span the entire Hilbert space.  In general, an $n$-blip excitation state localised at a position $x$, propagating in the $s$ direction and polarised in the $\lambda$ direction is given by
\begin{eqnarray}
	\label{blipstate2}
	\ket{n_{s\lambda}(x,t)} = \frac{a^\dagger_{s\lambda}(x,t)^n}{\sqrt{n!}}\, \ket{0} \, .
\end{eqnarray}

\subsubsection{The blip commutation relations} 
\label{222}

As we have seen above, our system may contain any number of identical blips.  The states that represent them, therefore, are unchanged when an exchange of blips takes place, and the blip creation and annihilation operators must each commute amongst themselves.  Hence, we assume in the following that
\begin{eqnarray}
	\label{blipcommutator1}
	\left[a_{s\lambda}(x,t),a_{s'\lambda'}(x',t')\right] = \left[a^\dagger_{s\lambda}(x,t), a^\dagger_{s'\lambda'}(x',t')\right] = 0 \, .
\end{eqnarray}
Using the definition of a single-excitation blip state in Eq.~(\ref{blipstate1}), it is possible to show that the commutation relation between blip creation and annihilation operators is identical to the inner product between two single-blip states, and does not necessarily vanish:
\begin{eqnarray}
	\braket{1_{s\lambda}(x,t)|1_{s'\lambda'}(x',t')} = \bra{0}\left[a_{s\lambda}(x,t), a^\dagger_{s'\lambda'}(x',t')\right]\ket{0}
	=\left[a_{s\lambda}(x,t), a^\dagger_{s'\lambda'}(x',t')\right] \, .
\end{eqnarray}
In order for blips to be strictly localised in space, single-blip states localised at different positions at a given time must be orthogonal to one another. A state localised at one position then has no chance of being found anywhere else. Polarisation, we know, is a measurable quantity; therefore, states with different polarisations can be distinguished. As was discussed in the introduction, here we must treat states parametrised by different $s$ as distinguishable too: blip states characterised by different $s$ and $\lambda$ must also be orthogonal.  Hence, in the following we demand that 
\begin{eqnarray} 
	\label{blipproduct1}
	\langle 1_{s\lambda} (x,t) |1_{s'\lambda'} (x',t') \rangle \nonumber 
	&=& \langle 1_{s\lambda} (x-sct,0) |1_{s'\lambda'} (x'-s'ct',0) \rangle \nonumber \\
	&=&\delta_s((x-sct)-(x'-s'ct')) \, \delta_{s,s'} \, \delta_{\lambda,\lambda'}
\end{eqnarray}
with $\delta_s(x-x')$ given by
\begin{eqnarray} 
	\label{commutator1}
	\delta_s(x-x') = \int_{-\infty}^{\infty} {\text{d}k \over 2\pi} \, {\rm e}^{{\rm i}sk(x-x')} \, .
\end{eqnarray}
The above inner product is a good choice because it is strictly positive, translation independent, symmetric with respect to the position of the blips and real valued.  There is also a unit probability of finding the blip within $(-\infty,\infty)$.  Hence, we find, at equal times, that
\begin{eqnarray} 
	\label{blipcommutator2}
	\left[ a_{s\lambda} (x), a^\dagger_{s'\lambda'} (x') \right] = \delta_s(x-x')  \, \delta_{s,s'} \, \delta_{\lambda,\lambda'} \, . 
\end{eqnarray}
This is the bosonic commutation relation which is expected to hold for all photonic particles.

\subsubsection{A fundamental equation of motion} 

Typically, the dynamics of photon states are calculated using Heisenberg's equation of motion. To obtain this equation, we need to know the Hamiltonian of the EM field which we do not have yet. When single-photon states are represented in the basis of energy eigenstates, obtaining a Hamiltonian is a simple process. However, the blip states have a well defined position in space and time, and Heisenberg's uncertainty relation therefore tells tell us that their momenta and energies are completely unknown.  Consequently, at this stage we cannot follow the usual approach to obtain an equation of motion. Fortunately, we may determine the dynamics of blip states and blip operators by another method.

Blip states represent the localised excitations of the EM field that propagate at the speed of light.  This assumption places a constraint on the expectation values of the EM fields at different times which then ensures propagation at a constant speed. This constraint is given by $\braket{a_{s\lambda}(x,t)} = \braket{a_{s\lambda}(x-sct,0)}$.  Since this relation holds for any time-independent state we can deduce the relation
\begin{eqnarray}
	\label{blipconstraint1}
	a_{s\lambda}(x,t) = a_{s\lambda}(x-sct,0) \, .
\end{eqnarray} 
This equality asserts that, when allowed to propagate freely, a blip state placed at a position $x$ at time $t=0$ will be found at a position $x+sct$ at the later time $t$.  Rather than invoking Heisenberg's equation, we are able to determine the equation of motion for a blip state using the above condition. By taking the time derivative of the blip state in Eq.~(\ref{blipconstraint1}) we may show that
\begin{eqnarray}
	\label{blipmotion1}
	\frac{\partial }{\partial t}a_{s\lambda}(x,t) = -sc\frac{\partial }{\partial x}a_{s\lambda}(x,t) \, .
\end{eqnarray}
This is the fundamental equation of motion for blip states.  

\subsection{Observables in the position representation} \label{Sec2.2}

\subsubsection{Field observables}
\label{Sec:1Dposfields}

In Section \ref{Sec:blipstates}, we constructed a new Hilbert space spanned by the blip number states (cf.~Eq.~(\ref{blipstate2})). Next we shall obtain a set of expressions for the (complex) operators ${\boldsymbol E}(x,t)$ and ${\boldsymbol B}(x,t)$, and the energy observable $H_\text{energy}(t)$.  As was shown in Section \ref{Sec:classical1}, the classical solutions of Maxwell's equations in one dimension obey the blip equation of motion in Eq.~(\ref{blipmotion1}).  Consequently, like the blips, the solutions of Maxwell's equations in a one-dimensional homogeneous medium are wave packets which travel at the relevant speed of light along the $x$ axis.  Hence,  in the following we postulate that the observables of the complex vectors ${\boldsymbol E}(x,t)$ and ${\boldsymbol B}(x,t)$ are given by
\begin{eqnarray} 
	\label{fieldobservables1}
	{\boldsymbol E}(x,t) &=& \sum_{s= \pm 1} c \, \left( \mathcal{R} \left[a_{sH}\right](x,t)\,\hat{\textbf{y}} + \mathcal{R}\left[a_{sV}\right](x,t)\,\hat{\textbf{z}} \right) \, , \nonumber \\
	{\boldsymbol B}(x,t) &=& \sum_{s= \pm 1} s \, \left( - \mathcal{R}\left[a_{sV}\right](x,t)\,\hat{\textbf{y}} + \mathcal{R} \left[a_{sH}\right](x,t)\,\hat{\textbf{z}} \right) \, .
\end{eqnarray}
The above operators are non-Hermitian, and their expectation values are complex by construction.  We assume here that the actual field expectation values are given by the real combination $(\boldsymbol{O}+\boldsymbol{O}^\dagger)/2$.  We shall resume this convention throughout the rest of this paper unless we make specific mention otherwise. The superoperator $\mathcal{R}$ has been added to the above equation to guarantee their validity. As we shall see below, in position space, the field observables cannot be written as linear superpositions of field annihilation and creation operators, despite this being the case in momentum space.

\subsubsection{The regularisation operator}

Next let us have a closer look at the properties of this operator $\mathcal{R}$ in Eq.~(\ref{fieldobservables1}), which we shall refer to as the regularisation operator. 
First we notice that it cannot depend on $x$ and $t$, and it must be symmetric and translation-invariant. It also cannot depend on $s$ and $\lambda$ if we aim for a description of light in which the same physics applies for field vectors of any orientation and travelling in any direction. Hence $\mathcal{R}$ can be understood in a distributional sense such that
\begin{eqnarray}
	\label{Rsuperposition1}
	\mathcal{R}\left[a_{s\lambda}\right](x,t) = \int_{-\infty}^{\infty}\text{d}x'\, R(x-x')\,a_{s\lambda}(x',t) \, ,
\end{eqnarray}
where $R(x-x')$ denotes a function that depends only on the distance between $x$ and $x'$. Its derivation will be provided later on in Section \ref{Sec:fieldcomparison}. Similarly, $\mathcal{R}\left[a^\dagger_{s\lambda}\right]$ is defined such that
\begin{eqnarray}
	\label{Rsuperposition2}
	\mathcal{R}\left[a^\dagger_{s\lambda}\right](x,t) = \int_{-\infty}^{\infty}\text{d}x'\, R^*(x-x')\,a^\dagger_{s\lambda}(x',t) \, .
\end{eqnarray}
The purpose of this distribution is to relate the measurable field observables to a local and causal particle in a possibly non-local way.  As the $\mathcal{R}$ superoperator is translation invariant, both the field observables and the blip states propagate at the speed of light and satisfy the equation of motion given in Eq.~(\ref{blipmotion1}).  It is then possible to show that the field observables in Eq.~(\ref{fieldobservables1}), and therefore also their expectation values, obey the free-space Maxwell equations.  Symmetry of $\mathcal{R}(x-x')$ is assumed due to the symmetry of the blip states; although their direction of propagation will be reversed, a parity transformation will only displace a blip and not change its shape.  It is therefore important that the electric and magnetic fields associated with a single-blip state are also only translated under a parity transformation.  

\subsubsection{The energy observable}

The magnitude of the regularisation operator also determines the energy expectation value of each individual blip state. To determine the energy observable in terms of the blip creation and annihilation operators, we substitute the field observables, Eq.~(\ref{fieldobservables1}) in this case, into the expression for the classical energy Eq.~(\ref{classicalenergy1}).  The resulting expression is
\begin{eqnarray} 
	\label{Heng1}
	H_{\text{energy}}(t)
	= \sum_{s= \pm 1} \sum_{\lambda = {\sf H},{\sf V}}  \int_{-\infty}^{\infty} \text{d}x \, {\varepsilon_0 c^2 A \over 4} \left( \mathcal{R}\left[a_{s\lambda}\right](x,t) + {\rm H.c.} \right)^2\, .
\end{eqnarray}
Thus, $\mathcal{R}$ determines the energy of a blip state.  Most importantly, because of the quadratic form of this observable, energy expectation values are always positive, as they are in classical electrodynamics.

\subsection{The dynamical Hamiltonian}
\label{Sec:dynHamiltonian1}

In this final subsection we would like to show that the equation of motion for a blip operator can be written as a Schr{\"o}dinger equation. More specifically, we would like to show that the field observables, ${\boldsymbol O}(x,t)$, evolve in the Heisenberg picture according to Heisenberg's equation of motion,
\begin{eqnarray}
	\label{Heisenberg's equation}
	{\partial \over \partial t} \, {\boldsymbol O}(x,t) = -\frac{\rm i}{\hbar}\left[ {\boldsymbol O}(x,t), H_\text{dyn}(t) \right]\,,
\end{eqnarray}
for some dynamical Hamiltonian $H_\text{dyn}(t)$. To deduce this Hamiltonian, we initially consider Heisenberg's equation of motion for the operator $a_{s\lambda} (x,t)$, which is given by
\begin{eqnarray} \label{blipmotion2}
	{\partial \over \partial t} \, a_{s \lambda}(x,t) = - {{\rm i} \over \hbar} \left[ a_{s\lambda}(x,t), H_\text{dyn}(t) \right] \, .
\end{eqnarray}
What is interesting about the blip annihilation operators is that their equation of motion is already known. It can be found in Eq.~(\ref{blipconstraint1}) which implies Eq.~(\ref{blipmotion1}). Using these equations allows us to replace the time derivative on the left-hand side of Eq.~(\ref{blipmotion2}) with a space derivative and to write Heisenberg's equation as
\begin{eqnarray} \label{blipmotion3}
	\frac{\partial }{\partial x} \, a_{s\lambda}(x,t) = {{\rm i}s \over \hbar c} \left[a_{s\lambda}(x,t), H_{\text{dyn}}(t) \right] \, .
\end{eqnarray}
The above equation of motion suggests that the dynamical Hamiltonian affects the position, but not the time coordinate, of $a_{s\lambda}(x,t)$. This is not surprising: the purpose of $H_\text{dyn}(t)$ is to propagate wave packets at the speed of light along the $x$ axis. As the generator of such dynamics, the Hamiltonian must continuously annihilate blips at their respective positions $x'$ while simultaneously replacing them with excitations of equal amplitudes at nearby positions $x''$ different from $x'$.  

Taking this into account, we may construct a general exchange Hamiltonian for blips at different locations:
\begin{eqnarray} \label{Hdyn1}
	H_{\text{dyn}}(t) = \sum_{s=\pm 1} \sum_{\lambda = {\sf H}, {\sf V}} \int_{-\infty}^{\infty}\text{d}x' \int_{-\infty}^{\infty}\text{d}x'' \, \hbar sc \, f_{s \lambda}(x'',x') \, a^\dagger_{s\lambda}(x'',t) a_{s\lambda}(x',t) \, ,
\end{eqnarray}
where the factor $\hbar sc$ has been added for later convenience and where $f_{s\lambda}(x'',x')$ is a complex function left to be determined.  By substituting the Hamiltonian in Eq.~(\ref{Hdyn1}) into our modified Heisenberg's equation, Eq.~(\ref{blipmotion3}), one finds that
\begin{eqnarray} \label{blipmotion4}
	\frac{\partial }{\partial x} \, a_{s\lambda}(x,t) = {\rm i} \int_{-\infty}^{\infty}\text{d}x' \, f_{s \lambda}(x-x') \, a_{s\lambda}(x',t) \, .
\end{eqnarray} 
From this equation, one is able to verify at once that
\begin{eqnarray}
	\label{deltaderivative}
	f_{s \lambda}(x-x') = \int_{-\infty}^{\infty} {\text{d}k \over 2\pi} \, sk \, {\rm e}^{{\rm i}sk(x-x')} 
	= - {\rm i} \, {\partial \over \partial x} \delta_s (x-x') 
	= - {\rm i} \, \delta_s'(x-x') \, .
\end{eqnarray}
Here $\delta_{s}'(x-x')$ denotes the derivative of the delta function $\delta_s(x-x')$, which we introduced previously in Eq.~(\ref{commutator1}), with respect to $x$.  

Overall, the dynamical Hamiltonian $H_\text{dyn}(t)$ in Eq.~(\ref{Hdyn1}) equals
\begin{eqnarray}
	\label{Hdyn2}
	H_{\text{dyn}}(t) = \sum_{s=\pm 1} \sum_{\lambda = {\sf H}, {\sf V}} \int_{-\infty}^{\infty}\text{d}x' \int_{-\infty}^{\infty}\text{d}x''  \int_{-\infty}^\infty \text{d} k \, {\hbar c k \over 2 \pi} \, {\rm e}^{{\rm i}sk (x''-x')} \, a^\dagger_{s\lambda}(x'',t) \,a_{s\lambda}(x',t) \, .
\end{eqnarray}
This Hamiltonian is Hermitian, and therefore a generator of unitary dynamics.  It also satisfies a number of relevant properties.  For example, $f_{s\lambda}(x-x')$ is antisymmetric under an exchange of $x$ and $x'$. This means that a state that propagates from $x$ to $x'$ will only propagate from $x'$ to $x$ if either $s$ is reversed or time is reversed. It also means that a blip with a well-defined direction of propagation $s$ cannot be replaced by another at the same position. Moreover, $f_{s\lambda}(x-x')$ is translation invariant, which means that blips propagate identically at all positions, as would be expected in free space.

Unlike the energy observable in Eq.~(\ref{Heng1}), the dynamical Hamiltonian in Eq.~(\ref{Hdyn2}) has both positive and negative eigenvalues.  In the standard quantum field theory, the dynamical Hamiltonian and energy observable are equal; however, as pointed out already in the Introduction, this no longer applies to the quantised EM field. The reason that the two are not the same will be discussed in section \ref{Sec:dynHamiltonian2}, but at present it is necessary for us to check that the energy of the system is conserved.  According to the Heisenberg equation, the observable for a conserved quantity commutes with the dynamical Hamiltonian.  When the dynamical Hamiltonian and energy observable coincide, this property it ensured automatically because all observables commute with themselves.  We must now check that  
\begin{eqnarray}
	\label{energyconservation}
	\left[ H_\text{energy}(t), H_\text{dyn}(t) \right] = 0 \, . 
\end{eqnarray} 
Owing to the symmetry of the distribution $R(x-x')$ and the antisymmetry of $f_{s\lambda}(x,x')$, one can show that this commutator indeed vanishes.

\section{Quantisation in the momentum representation}
\label{Sec:momentum}

In classical electrodynamics, the momentum and position representations of the EM field complement each other well, and may be used interchangeably for our convenience. For example, we often describe light scattering experiments using Maxwell's equations, which involve only local field amplitudes. In such situations, it might be best to use the position space representation when modelling the quantised EM field. In other situations, classical electrodynamics introduces optical Green's functions and decomposes the EM field into monochromatic waves to predict general optical properties \cite{Barcellona}. This is when it might be more convenient to consider a momentum-space representation. In addition to providing us with a more complete formulation of the quantised EM field, by investigating the momentum representation, we shall be able to examine more closely the relationship between the description of the previous section and the standard quantum optics description of the quantised EM field \cite{Bennett}.

\subsection{Quantisation in momentum space}

\subsubsection{Photon states}

From classical electrodynamics we know that the set of travelling waves with wave numbers $k \in (-\infty,\infty)$ and two different polarisations $\lambda$ provide a complete set of solutions of Maxwell's equations for light propagation in one dimension.  Usually, in momentum space, we therefore describe the quantised EM field with the help of annihilation operators $a_{k \lambda}$ with $k$ and $\lambda$ referring to the corresponding photon wave number and polarisation respectively \cite{Bennett}. However, when applying a Fourier transform to the blip annihilation operators $a_{s \lambda} (x,t)$ introduced in Section \ref{Sec2.2}, we obtain annihilation operators $\tilde a_{s \lambda} (k,t)$. In the following we assume that 
\begin{eqnarray}
	\label{FieldFT1}
	\tilde{a}_{s\lambda}(k,t) = \int_{-\infty}^{\infty}\frac{\text{d}x}{\sqrt{2\pi}}\, {\rm e}^{-{\rm i}skx} \, a_{s\lambda}(x,t) 
\end{eqnarray}
represent the annihilation operators of monochromatic photons with the inverse transformation 
\begin{eqnarray}
	\label{FieldFT2}
	a_{s\lambda}(x,t) = \int_{-\infty}^{\infty}\frac{\text{d}x}{\sqrt{2\pi}}\, {\rm e}^{{\rm i}skx} \, \tilde a_{s\lambda}(k,t) \, .
\end{eqnarray} 
Notice that the $\tilde{a}_{s\lambda}(k,t)$ operators are labelled by three parameters, $s = \pm 1$, $k \in (-\infty,\infty)$ and $\lambda = {\sf H}, {\sf V}$. As we shall see below and as pointed out already in Ref.~\cite{Jake}, the position space representation of the EM field which we introduce in this paper leads to a doubling of its Hilbert space compared to the standard description \cite{Bennett}. The need for this doubling is not surprising. While the electric field amplitudes of classical electrodynamics are always real, the wave functions of photons have complex coefficients and hence their description requires an additional degree of freedom. The reasons for including the factor $s$ in the exponents of the Fourier transforms in Eqs.~(\ref{FieldFT1}) and (\ref{FieldFT2}) will become more obvious below. For example, by choosing the signs of the exponents in the above way, we ensure that left- and right-moving wave packets have a different Fourier decomposition in momentum space, even when their position space coefficients are the same and the wave packets are of the same shape. 

Usually, for light propagating along the $x$ axis, the wave number $k$ is the $x$ component of the wave vector which is oriented in the direction of propagation. Moreover, its magnitude $|k|$ relates to the angular frequency $\omega$ through $\omega = c|k|$. In this section, we adopt the convention that the $x$ component of the wave vector is given by $sk$ \cite{Jake}. The parameter $s$, as before, indicates the direction of propagation.  We include it in the definition of the wave vector so that an inversion of the direction of propagation, i.e.~replacing $s$ by $-s$, reverses the wave vector, as a change in direction usually does.  In this way, $k$ loses its traditional interpretation. Nevertheless, to ensure that the Fourier transform in Eq.~(\ref{FieldFT2}) is invertible, we must assume that $k$ can take any real value \cite{Howell}. 

In the following, we refer to the energy quanta obtained when applying $\tilde a^\dagger_{s \lambda}(k,t)$ to the vacuum state as photons, but one should notice that these photons are not exactly the same as the photons in the standard description of the EM field \cite{Bennett}. For example, the coherent states of the operators $\tilde{a}_{-1\lambda}(-k,t)$ and $\tilde{a}_{+1\lambda}(k,t)$ may have exactly the same electric field expectation values but superposing $\tilde{a}_{-1\lambda}(-k,t)$ and $\tilde{a}_{+1\lambda}(k,t)$ photons creates left- and right-moving wave packets respectively. Both operators $\tilde{a}_{-1\lambda}(-k,t)$ and $\tilde{a}_{+1\lambda}(k,t)$ are associated with {\em different} dynamics. As pointed out already in Section \ref{sec1}, although the standard $a_{k \lambda}$ operators can be used to create wave packets with electric field expectation values of any shape, they cannot create localised wave packets of arbitrary shape which move at the speed of light, i.e.~without dispersion, in a well-defined direction.

As in position space, the total Hilbert space of the quantised EM field can be obtained by applying creation operators repeatedly to the vacuum state. Because of the linearity of the above equations, the vacuum state $\ket{0}$ is still the state for which  
\begin{eqnarray}
	\label{annihilate2}
	\tilde a_{s \lambda}(k,t) \, \ket{0} = 0 
\end{eqnarray}
for all $s$, $\lambda$, $k$ and $t$. As in Section \ref{Sec:blipstates}, by applying a creation operator $\tilde a^\dagger_{s \lambda}(k,t)$ to the vacuum state once we obtain a single photon state
\begin{eqnarray} 
	\label{Photonstate1xx}
	\ket{\tilde 1_{s\lambda} (k,t)} = \tilde a^{\dagger}_{s\lambda}(k, t) \ket{0} \, .
\end{eqnarray}
As in Section \ref{222}, by applying a creation operator $\tilde a^{\dagger}_{s\lambda}(k, t)$ to the vacuum state an arbitrary number of times, we may generate states containing any number of photons. For example,
\begin{eqnarray} 
	\label{41}
	\ket{\tilde n_{s\lambda} (k,t)} = {\tilde a^{\dagger}_{s\lambda}(k, t)^n \over \sqrt{n}} \ket{0} 
\end{eqnarray}
is a state with exactly $n$ excitations in the $(s,k,\lambda,t)$ photon mode.

\subsubsection{The photon commutation relations}

Next we determine a set of commutation relations for the annihilation and creation operators $\tilde a_{s \lambda}(k,t)$ and $\tilde a^\dagger_{s \lambda}(k,t)$. Because photons are bosons, multi-photon states are unchanged when two or more photons are exchanged. This leads us to the typical commutation relation for bosonic particles:
\begin{eqnarray}
	\label{photoncommutator1}
	\left[ \tilde a_{s \lambda}(k,t), \tilde a_{s'\lambda'}(k',t')\right] = \left[ \tilde a^\dagger_{s \lambda}(k,t), \tilde a^\dagger_{s'\lambda'}(k',t')\right] = 0  \, .
	\end{eqnarray}
These relations are analogous to and in agreement with Eq.~(\ref{blipcommutator1}). If annihilation operators\textemdash and creation operators respectively \textemdash commute with each other in position space, the same must hold in momentum space, since both are connected via Fourier transforms. Substituting Eq.~(\ref{FieldFT1}) into the blip commutator relations in Eq.~(\ref{blipcommutator2}) and performing the resulting integrals, one can moreover show that 
\begin{eqnarray}
	\label{photoncommutator3}
	\left[ a_{s\lambda}(k,t), a^\dagger_{s'\lambda'}(k',t)\right] = \delta(k-k') \, \delta_{s,s'} \, \delta_{\lambda,\lambda'} \, .
\end{eqnarray}
Hence the single-photon states in Eq.~(\ref{Photonstate1xx}) can be shown to be orthogonal to one another;
\begin{eqnarray} 
	\label{photoncommutator2}
	\braket{\tilde 1_{s\lambda}(k,t)|\tilde 1_{s'\lambda'}(k',t)} = \braket{0|\left[ a_{s\lambda} (k,t), a^\dagger_{s'\lambda'} (k',t) \right]|0} 
	= \delta(k-k') \, \delta_{s,s'} \, \delta_{\lambda,\lambda'} \, ,
\end{eqnarray}
as they are in standard quantum electrodynamics. As one would expect, the photons that we consider in this section are bosonic particles. 

\subsubsection{The fundamental equation of motion} 

We may now determine the time-dependence of the photon creation and annihilation operators by substituting the Fourier transform in Eq.~(\ref{FieldFT1}) into the equation of motion of the blip annihilation operators $a_{s\lambda}(x,t)$ which can be found in Eq.~(\ref{blipconstraint1}). Doing so, one is then able to verify that the time-dependent photon annihilation operators evolve such that 
\begin{eqnarray}
	\label{photonevolution1}
	\tilde a_{s \lambda}(k,t) = {\rm e}^{-{\rm i} ck t} \, \tilde a_{s \lambda}(k,0) \, .
\end{eqnarray} 
This equation is the usual equation of motion of the quantised EM field in momentum space and shows that photons oscillate with an angular frequency $ck$. However, since $k$ now varies between $- \infty$ and $+ \infty$, the angluar frequency $ck$ can be positive and negative. This is important because, without considering the full range of frequencies, the Fourier transform in Eq.~(\ref{FieldFT2}) would not have an inverse transformation (cf.~Eq.~(\ref{FieldFT1})). As we have illustrated in the Introduction and as we have seen in the previous section, this is not a problem. As we shall see below, photon states always have positive energy expectation values. This is possible since the energy observable $H_\text{energy}(t)$ of the quantised EM field no longer coincides with the generator of its dynamics.

\subsection{Observables in the momentum representation}
\label{Sec:1Dmomfields}

\subsubsection{Field observables}

As mentioned above, it is often convenient to express the position-dependent electric and magnetic fields in their Fourier representations. In the following, we denote them $\widetilde {\boldsymbol{E}}(k,t)$ and $\widetilde {\boldsymbol{B}}(k,t)$ respectively. Like the fields themselves, they are 3-vector valued, and, in this case, they are parametrised by a time $t$ and a real wave number $k$. As we have seen in Section \ref{Sec:position}, each individual component of the electric and magnetic field vectors are linear superpositions of travelling waves with a well-defined direction of propagation $s$ and polarisation $\lambda$. Since the Fourier transform is linear, the same applies to their Fourier components and 
\begin{eqnarray}
\widetilde {\boldsymbol{O}}(k,t) &=& \sum_{s=\pm 1} \sum_{\lambda = {\sf H}, {\sf V}} \widetilde {\boldsymbol{O}}{s\lambda}(k,t)
\end{eqnarray} 
with $\boldsymbol{O} = \boldsymbol{E},\boldsymbol{B}$. The different components in this equation have different Fourier modes and
\begin{eqnarray}
	\boldsymbol{O}_{s\lambda}(x,t) &=& \int_{-\infty}^{\infty}\frac{\text{d}k}{\sqrt{2\pi}}\,{\rm e}^{{\rm i}skx} \, \widetilde{\boldsymbol{O}}_{s\lambda}(k,t) \, , \nonumber \\
	\widetilde{\boldsymbol{O}}_{s\lambda}(k,t) &=& \int_{-\infty}^{\infty}\frac{\text{d}x}{\sqrt{2\pi}}\, {\rm e}^{-{\rm i}skx} \, \boldsymbol{O}_{s\lambda}(x,t)
\end{eqnarray} 
for all transformation to be consistent with Eqs.~(\ref{FieldFT1}) and (\ref{FieldFT2}). Keeping this in mind and applying the respective Fourier transform to the differential equation in Eq.~(\ref{11}), one may show that the electric field $\widetilde {\boldsymbol{E}}_{s\lambda}(k,t)$ satisfies the first-order differential equation
\begin{eqnarray}
	\label{waveequation2}
	\left({\rm i}ck + \frac{\partial}{\partial t}\right)\widetilde{\boldsymbol{E}}_{s\lambda}(k,t) = 0 \, .
\end{eqnarray}
In analogy to standard quantisations of the EM field, in which the system is described as a collection of simple harmonic oscillators \cite{Heitler}, one can show that the complex electric and magnetic field observables $\widetilde {\boldsymbol{E}}(k,t)$ and $\widetilde {\boldsymbol{B}}(k,t)$ are linear combinations of photon annihilation operators of the form
\begin{eqnarray} 
	\label{FTfieldcomponents2}
	\widetilde{\boldsymbol{E}}(k,t) &=& \sum_{s=\pm 1} c\, \Omega(k)\, {\rm e}^{{\rm i}\varphi(k)} \left( \tilde a_{s {\sf H}}(k,t) \, \hat{\mathbf{y}} +  \tilde a_{s {\sf V}}(k,t) \, \hat{\mathbf{z}} \right)  \, , \nonumber\\
	\widetilde{\boldsymbol{B}}(k,t) &=& \sum_{s=\pm 1} s\, \Omega(k) \, {\rm e}^{{\rm i}\varphi(k)} \left( - \tilde a_{s {\sf V}}(k,t) \, \hat{\mathbf{y}} + \tilde a_{s {\sf H}}(k,t) \, \hat{\mathbf{z}} \right) \, .
\end{eqnarray}
Here $\Omega(k)$ is a $k$-dependent numerical factor. By introducing the $k$-dependent phases $\varphi(k)$, we may assume that $\Omega(k)$ is real. In the standard approach, the above equations can be justified by noticing that the corresponding energy observable $H_\text{energy}(t)$ must take the form of a harmonic oscillator Hamiltonian \cite{Bennett}. Here it can be justified by substituting Eq.~(\ref{photonevolution1}) for the dynamics of the $\tilde a_{s \lambda}(k,t)$ operators into the above equations and checking that Eq.~(\ref{waveequation2}) holds. 

\subsubsection{Normalisation of electric and magnetic field amplitudes}
\label{Sec:fieldamplitudes}

However, the fundamental equation of motion, Eq.~(\ref{photonevolution1}), cannot be used to determine $\Omega(k)\, {\rm e}^{{\rm i}\varphi(k)}$. The factor $\Omega(k)$ is a function of $k$ which determines the field amplitudes and therefore also the energy of a single photon in the $(k,s,\lambda)$ mode. Due to the homogeneity of the EM field, we can safely assume that $\Omega(k)$ does not have any dependence on the parameters $s$ or $\lambda$.  For the time being we shall not specify $\Omega(k)$ any further, but we shall return to this function in Section \ref{Sec:Lorentzcovariance}.

\subsubsection{The energy observable}
\label{Sec:momenergyobservable}

For completeness we now also derive the energy observable of the quantised EM field, $H_{\text{energy}}(t)$, in the momentum representation. Taking again the expression for the classical field energy, Eq.~(\ref{classicalenergy1}), as our starting point and substituting in for the classical fields the position-dependent field observables in their Fourier representations, one finds that
\begin{eqnarray}
	\label{Heng3}
	H_{\text{energy}}(t) = \sum_{s= \pm 1} \sum_{\lambda = {\sf H},{\sf V}} \int_{-\infty}^{\infty}\text{d}k \,  {\varepsilon_0 c^2 A \over 4} \, \left\| \Omega(k)\, 
	{\rm e}^{{\rm i} \varphi(k)} \, \tilde a_{s \lambda}(k,t) + \Omega^*(-k)\, {\rm e}^{-{\rm i}\varphi(-k)} \, a^\dagger_{s\lambda}(-k,t) \right\|^2 .
\end{eqnarray}
The expectation values of the above energy observable are again positive. This is guaranteed by the modulus in the integrand. Moreover, we notice that the above energy observable has an explicit dependence on the choice of phase $\varphi(k)$.  When we restrict this theory to positive wave numbers only, this dependence disappears; however, we cannot make that assumption here.  The absolute phase of a field is not observable, and therefore should not influence the energy of the field. To remove this unwanted dependence we must impose the following condition: 
\begin{eqnarray}
\varphi(k) = -\varphi(-k) \, .  
\end{eqnarray}
One can see that the phase gained by evolving the system in time is of this form. By substituting into this expression the explicit time dependence of the photon operators given in Eq.~(\ref{photonevolution1}), one can verify that the energy observable is time-independent.

\subsection{The dynamical Hamiltonian}
\label{Sec:dynHamiltonian2}

Using Eq.~(\ref{photonevolution1}) one can show that the $n$-photon states $\ket{\tilde n_{s\lambda}(k,t)}$ in Eq.~(\ref{41}) are, up to the accumulation of a time-dependent phase factor, invariant under the dynamical evolution of the EM field. Hence they must be eigenstates of the dynamical Hamiltonian $H_{\rm dyn}(t)$.  The eigenvalue corresponding to the state $\ket{\tilde n_{s\lambda}(k,t)}$ is $n \hbar c k$. Given the bosonic commutation relation in Eq.~(\ref{photoncommutator3}), it is therefore straightforward to show that the dynamical Hamiltonian is given by
\begin{eqnarray}
	\label{dynamical Hamiltonian}
	H_{\text{dyn}} (t) = \sum_{s= \pm 1} \sum_{\lambda = {\sf H},{\sf V}} \int_{-\infty}^{\infty}\text{d}k \, \hbar ck \, \tilde a^\dagger_{s \lambda}(k,t)\, \tilde a_{s \lambda}(k,t) \, .
\end{eqnarray}
in momentum space. Using Eq.~(\ref{photonevolution1}), it is possible to show that the dynamical Hamiltonian is time independent. Moreover, using the Fourier transforms which we introduced at the beginning of this section (cf.~Eqs.~(\ref{FieldFT1}) and (\ref{FieldFT2})), one can check that this Hamiltonian is the same as the dynamical Hamiltonian which we obtained previously.

In section \ref{Sec:dynHamiltonian1}, we pointed out that the dynamical Hamiltonian for blip states had both positive and negative eigenvalues.  This ensures that the dynamics of localised light pulses is reversible: light moving to the left is indistinguishable from light moving to the right when time is reversed.  Here we have found that, if our system contains photons of negative $k$, then the dynamical Hamiltonian in this representation also possesses positive and negative eigenvalues.
In the momentum representation, the dynamical Hamiltonian, being diagonal, takes a much simpler form than the equivalent expression in the position representation (cf.~Eq.~(\ref{Hdyn2})).

\section{The importance of position and momentum representations}
\label{sec:relationship}

In the remainder of this paper let us emphasize that both the position and the momentum space quantisation approaches are important to obtaining a complete picture of the quantised EM field. For example, studying the EM field in position space has helped us to identify an otherwise hidden degree of freedom, namely the parameter $s$ which characterises the direction of propagation of wave packets of light. In Section \ref{Sec:position}, we determined the Hilbert space for the modelling of light propagation in one dimension. By solving Maxwell's equations, we were also able to derive sets of field observables in addition to constructing a dynamical Hamiltonian that describes the time-evolution of the system. However, we were not able to determine the regularisation operator $\mathcal{R}$ which establishes a relationship between the local blip annihilation operators and the electric and magnetic field observables.  

Although both the position and the momentum descriptions that we present here are essentially equivalent, it easier to determine the normalisation factors of electric and magnetic field observables in momentum space. In the following, we determine these normalisation factors. The Fourier transforms in Eqs.~(\ref{FieldFT1}) and (\ref{FieldFT2}) allow us to alternate freely between the position and momentum representations at our pleasure. Once we have identified $\Omega(k)$, we can then draw conclusions about the effect of the position space regularisation operator $\mathcal{R}$. Finally, in this section, we have a closer look at the relationship between our approach and the standard description of the quantised EM field in momentum space. It is shown that the standard approach emerges when we restrict the photon annihilation operators that we consider in this paper to a certain subset. 

\subsection{Lorentz covariance}
\label{Sec:Lorentzcovariance}

In Section \ref{Sec:1Dmomfields}, we were able to define the electric and magnetic field observables only up to a $k$-dependent factor $\Omega(k)$ which was shown to be directly related to the energy of a photon. One way to determine this factor is therefore to presume the energy of a photon and to work backwards.  Another method is to ensure that the electric and magnetic field observables transform correctly under the proper orthochronous Lorentz transformations.  In the following we shall follow this latter approach.  

For excitations restricted to propagate in one dimension, the possible transformations, denoted by the greek letter $\Lambda$, are translations in $x$ and $t$, and rotations about and boosts along the $x$ axis.  The inner product between two states is a Lorentz scalar, and is therefore unchanged by any of the transformations above.  Naturally, these changes of reference frame induce a unitary operation on a state, which we shall denote $U(\Lambda)$.  Let us consider initially such transformations on a normalised single-photon state $\ket{\psi}$.  If we assume that the vacuum state is invariant under Lorentz transformations, a transformed single-photon state is given by
\begin{eqnarray}
	U(\Lambda)\ket{\psi} = \sum_{s\lambda}\int_{-\infty}^{\infty}\text{d}k\, \tilde \psi(k)\, U(\Lambda)\, \tilde a^\dagger_{s\lambda}(k)\,U^\dagger(\Lambda)\ket{0}.
\end{eqnarray}
Under a translation or a rotation, the photon creation operator gains only a phase factor \cite{Weinberg}.  The Lorentz boosts along the $x$ axis, however, involve a more interesting transformation.  Let us choose the particular transformation $\Lambda$ that causes a Doppler shift of the wave number $k$ to the new wave number $p$.  Let us further assume for simplicity that there is no rotation about the $x$ axis so that $s$ and $\lambda$ are unchanged.  By taking into account that the Lorentz-invariant measure for an integral over $k$ is given by $\text{d}k/|k|$, under a Lorentz boost, the normalised inner product $\braket{\psi|\psi}$ is form invariant only when
\begin{eqnarray}
	\label{photontransformation1}
	U(\Lambda)\, \tilde a^\dagger_{s\lambda}(k)\,U^\dagger(\Lambda) =  \sqrt{\left|\frac{p}{k}\right|}\, \tilde a_{s\lambda}(p).
\end{eqnarray}
In classical electromagnetism, the electric and magnetic field vectors in a Cartesian basis are the components of an antisymmetric rank-2 tensor given in the same basis, which have particular transformation properties when a change of reference frame is made, either by moving the field or moving ourselves (see, for example, \cite{Griffiths}).  We should expect that the expectation values of the field observables ${\boldsymbol E}(x,t)$ and ${\boldsymbol B}(x,t)$ transform in just the same way.  Using the transformation of photon operators given in Eq.~(\ref{photontransformation1}), we may show that the correct transformation occurs when 
\begin{eqnarray}
	\label{Omega1}
	\Omega(k) = \sqrt{|k|}\,\Omega_0.
\end{eqnarray}  
If we want the expectation values of the energy observable, $H_{\text{energy}}(t)$ in Eq.~(\ref{Heng3}), and of the dynamical Hamiltonian, $H_{\text{dyn}} (t)$ in Eq.~(\ref{dynamical Hamiltonian}), to be the same, at least in some cases, we must choose 
\begin{eqnarray}
	\label{photontransformation2}
	\left| \Omega_0 \right|^2 = {2 \hbar \over \varepsilon_0 c A} \, .
\end{eqnarray}
The latter equivalence only holds for states with positive values of $k$. In general, the above choice of $\Omega_0$ implies that the energy of a single photon in the $(s,k,\lambda)$ mode equals $\hbar c|k|$. 

\subsection{The regularisation operator revisited}
\label{Sec:fieldcomparison}

In the beginning of Section \ref{Sec:momentum}, we describe how to switch between the momentum and position representations of field observables with the help of Fourier transforms. Combining Eqs.~(\ref{fieldobservables1}) and (\ref{FieldFT1}) with $O_{s\lambda}(x,t) = a_{s\lambda}(x,t)$, one arrives at the expressions
\begin{eqnarray} 
	\label{fieldobservables4}
	{\boldsymbol E}(x,t) &=& \sum_{s= \pm 1} \int^{\infty}_{-\infty} {\text{d}k \over \sqrt{2 \pi}} \, c \, {\rm e}^{{\rm i}(skx + \varphi(k))} \, \left( \mathcal{R} \left[ \tilde a_{s{\sf H}}\right](k,t) \,\hat{\textbf{y}} + \mathcal{R}\left[ \tilde a_{s{\sf V}}\right](k,t)\,\hat{\textbf{z}}\right) , \nonumber \\
	{\boldsymbol B}(x,t) &=& \sum_{s= \pm 1} \int^{\infty}_{-\infty} {\text{d}k \over \sqrt{2 \pi}} \, s \, {\rm e}^{{\rm i}(skx + \varphi(k))} \, \left( - \mathcal{R}\left[\tilde a_{s{\sf V}}\right](k,t)\,\hat{\textbf{y}} + \mathcal{R} \left[\tilde a_{s{\sf H}}\right](k,t) \,\hat{\textbf{z}} \right) \,.
\end{eqnarray}
Moreover, substituting the results in Eqs.~(\ref{Omega1}) and (\ref{photontransformation2}) into the Fourier transforms (cf. Eq.~(\ref{FieldFT1})) of the electric and magnetic field observables in Eq.~(\ref{FTfieldcomponents2}), we find that
\begin{eqnarray} \label{fieldobservables5}
	{\boldsymbol E}(x,t) &=& \sum_{s= \pm 1} \int^{\infty}_{-\infty} \text{d}k \, c \,\sqrt{\frac{\hbar|k|}{\varepsilon_0 \pi c A}} \, {\rm e}^{{\rm i}(skx+ \varphi(k))} \, \left(\tilde a_{s {\sf H}}(k,t) \, \hat{\textbf{y}} +  \tilde a_{s {\sf V}}(k,t) \, \hat{\textbf{z}} \right) \, ,  \nonumber \\
	{\boldsymbol B}(x,t) &=& \sum_{s= \pm 1} \int_{-\infty}^{\infty}\text{d}k \, s \,\sqrt{\frac{\hbar |k|}{\varepsilon_0 \pi c A}} \, {\rm e}^{{\rm i}(skx+\varphi(k))} \, \left( - \tilde a_{s {\sf V}}(k,t) \, \hat{\textbf{y}} + \tilde a_{s {\sf H}}(k,t) \, \hat{\textbf{z}}  \right) \, . 
\end{eqnarray}
A comparison of the above sets of equations enables us to determine the action of the superoperator $\mathcal{R}$ on the photon creation and annihilation operators in momentum space and to show that
\begin{eqnarray} 
	\label{regularise1}
	\mathcal{R} \left[ \tilde a_{s\lambda}\right] (k,t) = \sqrt{2 \hbar |k| \over \varepsilon_0 c A} \, \tilde a_{s \lambda}(k,t) \, .
\end{eqnarray}
In the momentum representation, in order to regularise the photon operators we multiply them by the $k$-dependent factor $(2\hbar|k|/\varepsilon_0 c A)^{1/2}$.  As expected from the symmetries of the quantised EM field, the superoperator $\mathcal{R}$ has no dependence on $x$, $t$, $s$ or $\lambda$.  

In Eq.~(\ref{fieldobservables4}), we assumed that the regularised blip operators, which by themselves are the Fourier transforms of the photon operators, are equal to the Fourier transforms of the regularised photon operators.  The action of the regularisation superoperator $\mathcal{R}$ on a blip annihilation operator can therefore be determined by Fourier transforming into the momentum representation, regularising the photon operators and performing the inverse Fourier transformation. One will find that the action of the regularisation operator on a blip annihilation operator is of the type given in Eq.~(\ref{Rsuperposition1}) where the function $R(x-x')$ is given by
\begin{eqnarray}
  	R(x-x') = \int_{-\infty}^{\infty}{\text{d}k \over 2\pi} \, \sqrt{\frac{2\hbar|k|}{\varepsilon_0 c A}} \, {\rm e}^{{\rm i}sk(x-x')} \, .
\end{eqnarray}
The above trick of introducing the superoperator $\mathcal{R}$ allows us to describe the quantised EM field in position space in terms of local bosonic blip operators without having to sacrifice the Lorentz covariance of the local electric and magnetic field observables. Importantly, however, this distribution is not local in the same way that the Dirac delta function is.  We can see this by evaluating the function $R(x-x')$ at values $x \neq x'$:
\begin{eqnarray}
	\label{explicitR}
	R(x-x') = -\sqrt{\frac{\hbar}{4\pi\varepsilon_0 cA}} \cdot \frac{1}{|x-x'|^{3/2}} \, .
\end{eqnarray}
This means that the field observables are not a simple linear superposition of blip operators defined at the same point.  However, since it is easier to work with bosonic annihilation and creation operators, we can perform all calculations in the Hilbert space created by the local bosonic operators.  More will be said on this feature in sections \ref{Sec:biorthogonal} and \ref{Sec:photonsdiscussion}.  As a final point, we may mention that, due to the equivalence of the field observables, the energy observable is also equal in both representations. 

\subsection{Comments on field and blip localisation}
\label{Sec:biorthogonal}

Alternatively, some authors might prefer to work with non-locally acting photon annihilation operators with a closer link to local field observables \cite{Hawton6,Jake2}. Such operators are given by
\begin{eqnarray} 
	\label{fieldexcitation1}
	A_{s \lambda}(x,t) = \mathcal{R}[a_{s \lambda}](x,t) 
\end{eqnarray}
and describe excitations that share the vacuum state with the blip excitations. The reason we differentiate between blip operators, $a_{s\lambda}(x,t)$, and the field excitations, $A_{s\lambda}(x,t)$, is that the blip operators possess a set of bosonic commutation relations with respect to the conventional inner product of quantum physics (cf. Eq.(\ref{blipcommutator2})). In contrast to this, the commutation relation of the $A_{s \lambda}(x,t) $ operators is given by
\begin{eqnarray}
	\label{fieldexcitationscommutator1}
	\left[ A_{s\lambda}(x,t), A^\dagger_{s'\lambda'}(x',t) \right] = \int^\infty_{-\infty} \text{d}k \, |k| \, {\rm e}^{{\rm i}sk(x-x')} \, \delta_{ss'} \,\delta_{\lambda\lambda'} \, . 
\end{eqnarray}
Nevertheless, as one can see for example from Eq.~(\ref{fieldobservables1}), the energy quanta associated with the $A_{s\lambda}(x,t)$ operators can be linked more easily to local electric and magnetic field amplitudes. Indeed, their expectation values have the units of $(\text{energy density})^{1/2}$ and not $(\text{probability density})^{1/2}$ as would be expected for a wave function compatible with the Born rule.

With respect to the conventional inner product of quantum physics, the single field excitations, $\ket{1^A_{s\lambda}(x,t)} = A^\dagger_{s\lambda}\ket{0}$, are not orthogonal to one another. Because of Eq.~(\ref{fieldexcitationscommutator1}), one can show that
\begin{eqnarray} 
	\label{bioproduct1}
	\braket{ 1^{(A)}_{s\lambda}(x,t) | 1^{(A)}_{s'\lambda'}(x',t)} = \braket{ 0 | \left[A_{s\lambda}(x,t) | A^\dagger_{s'\lambda'}(x',t)\right] |0 } \neq \delta(x-x') \, .
\end{eqnarray}
Although we shall not make use of it here, for completeness, it is worth mentioning here that it is possible to construct a Hilbert space in which the field excitations associated with the $A_{s\lambda}(x,t)$ operators can be treated as local. This is achieved by introducing a new inner product\textemdash labelled by superscript ${(A)} $\textemdash such that
\begin{eqnarray} 
	\label{bioproduct1}
	\braket{ 1^{(A)}_{s\lambda}(x,t) | 1^{(A)}_{s'\lambda'}(x',t) }^{(A)} = \delta(x-x') \, . 
\end{eqnarray}
Under this new inner product, the single-photon field excitations form an orthogonal basis of states in the position representation. However, this new inner product drastically alters
 the Hilbert space of the quantised EM field. For example, some previously Hermitian operators are now no longer Hermitian, whereas others\textemdash previously non-Hermitian\textemdash become Hermitian.  For this reason, this approach is known as ``biorthogonal" or ``pseudo-Hermitian" quantum mechanics. It is an interesting area of physics that has attracted a lot of attention in the field of local quantum electrodynamics \cite{Raymer,Hawton4,Brody,Hawton6}.  Although this is a very elegant way of restoring orthogonality, constructing a biorthogonal system introduces complexities that are by no means necessary to our understanding of the dynamics of local photons \cite{Jake2}.

\subsection{The relation to standard descriptions}

In this final subsection, we compare the description of the EM field in Section \ref{Sec:1Dmomfields} with the standard description of the quantised EM field in momentum space and ask which additional assumptions have to be made for the latter to emerge. Looking at Eqs.~(\ref{Heng3}) and (\ref{dynamical Hamiltonian})  when $\Omega(k)$ is given by Eq.~(\ref{Omega1}), we can see that the energy observable and the dynamical Hamiltonian coincide when the negative-frequency photons are excluded and we restrict ourselves to positive values of $k$. As one can check relatively easily, in this case, the real parts of the local field observables in Eq.~(\ref{FTfieldcomponents2}) reduce to their more usual expressions \cite{Bennett}. As we know, the positive-frequency photon states provide a complete description of the quantised EM field in the sense that they can be superposed to reproduce the right electric and magnetic field expectation values for wave packets of any shape. However, as we have seen in the introduction, they are not sufficient to generate the quantum versions of all possible solutions of Maxwell's equations, like highly-localised wave packets that remain localised \cite{Hegerfeldt3}. 

\section{Discussion}
\label{Sec:photonsdiscussion}

The results in this paper are based on the following basic aspects of classical physics which must hold simultaneously. Firstly, in one dimension, we can localise wave packets of light to arbitrarily small length scales, i.e.~at positions $x$. Secondly, measurements are constant on the light cones. In Section \ref{Sec:position}, we used these properties to construct a straightforward and natural description of  one-dimensional quantised EM fields in position space.  Our starting point is the assumption that we can generate local particles of light---so-called bosons localised in position (blips)---by applying bosonic creation operators $a^\dagger_{s \lambda}(x,t)$ to the vacuum state $|0 \rangle$. Using the above postulates, we then identified their Schr\"odinger equation and constructed electric and magnetic field observables ${\boldsymbol E}(x,t)$ and ${\boldsymbol B}(x,t)$ that are consistent with Maxwell's equations. The Lorentz covariance of these field operators is achieved with the help of a regularisation superoperator $\mathcal{R}$ (cf.~Eqs.~(\ref{fieldobservables1}), (\ref{Rsuperposition1}) and (\ref{explicitR})). Although the blips themselves are local, they carry non-local fields which can be felt in a region of space surrounding the blips. Moreover, blip states can be used to construct wave packets of light of any shape which remain local when travelling at the speed of light along the $x$ axis.

In addition, we asserted in this paper that the position and momentum representations of the free EM field in our theory are equivalent representations of the same physical system.  This expression of equivalence assumes the following three conditions:
\begin{enumerate}
	\item The momentum and position Hilbert spaces have the same vacuum state $|0 \rangle$.
	\item There is a linear, invertible transformation between the position and momentum space annihilation operators, $a_{s\lambda}(x,t)$ and $\tilde a_{s \lambda}(k,t)$, that preserves their bosonic commutation relations (cf.~Eqs.~(\ref{blipcommutator2}) and (\ref{photoncommutator3})).
	\item All observables of the quantised EM field are equal in either representation. 
\end{enumerate}
The third condition guarantees that the expectation values of observables are identical in both representations. It also guarantees that the position representation of the EM field is Lorentz covariant. This is indeed the case if the normalisation of electric and magnetic field observables is carried out as described in Section \ref{sec:relationship}.

Finally, by writing the blip excitations as superpositions of monochromatic photons, we have shown that our approach is consistent with the standard theory of the quantised EM field with the addition of countable negative-frequency states.  Previously, these states have been widely overlooked but the concept of adopting them to negate the consequences of Hegerfeldt's theorem \cite{Hegerfeldt3} is not new.  The idea of negative-frequency excitations has long been realised as important in a local description of light.  For instance, Allcock pointed out that negative frequency modes were necessary to define states that have a well-defined and measurable time of arrival \cite{Allcock,Allcock2,Allcock3}. It is clear from Property $2$ above that we can specify when a blip state will arrive at any given position.    Negative-frequency field solutions have also been considered in Refs.~\cite{Hawton5,Cook1,Cook2,Mandel, Hawton6, Fabio, Dickinson, Bostelmann, newHawton}.  In this paper, by introducing local particles of light with a given direction of propagation, we have clarified how these solutions arise naturally in a covariant quantised theory. In addition, we exposed some consequences of a theory containing these states, such as the difference between the energy observable $H_{\text{energy}}(t)$ and the generator for time translation, i.e.~the dynamical Hamiltonian $H_{\text{dyn}}(t)$ \cite{Jake}.

In classical electrodynamics, a local description of the EM field is often preferable to a non-local description. Similarly, we expect that the modelling of the quantised EM field in terms of blip states is often preferable to the standard description in terms of monochromatic photon states.  For example, the position space representation in Section \ref{Sec:position} should provide an extremely useful tool for modelling the quantised EM field in inhomogeneous dielectric media and on curved spacetimes \cite{Maybee}. Moreover, a local description is advantageous when modelling local light-matter and local light-light interactions. For example, in Ref.~\cite{Jake}, we used the blip annihilation operators $a_{s \lambda}(x,t)$ to construct locally-acting mirror Hamiltonians. Other potential applications of the results in this paper include providing novel insight into fundamental effects, like the Fermi problem, the Casimir effect and the Unruh effect, as well as the modelling of linear optics experiments with ultra-broadband photons \cite{exp,exp2,exp3,exp4}. \\[0.5cm]
{\bf Acknowledgments.} AB and JS acknowledge financial support from the Oxford Quantum Technology Hub NQIT (grant number EP/M013243/1). DH acknowledges financial support form the UK Engineering and Physical Sciences Research Council EPSRC (Award Ref. Nr. 2130171).
Moreover, we acknowledge many stimulating and helpful discussions with M. Basil Altaie.


\begin{thebibliography}{99}
	\bibitem{Pryce}  
	M. H. L. Pryce, 	Philos. Trans. R. Soc. A {\bf 195}, 62 (1948). 
	
	\bibitem{Wigner}
	T. D. Newton and E. P. Wigner, Rev. Mod. Phys. {\bf 21}, 400 (1949).

	\bibitem{Hawton7} 
	M. Hawton and V. Debi\`ere, J. Math. Phys. {\bf 60}, 052104 (2019).
	
	\bibitem{Jordan}
	T. F. Jordan, J. Math. Phys. \textbf{19}, 1382 (1978).
	
         \bibitem{Fleming1} 
	G. N. Fleming, Phys. Rev. \textbf{137}, B188 (1965).
	
	\bibitem{Fleming3} 
	G. N. Fleming, Phys. Rev. \textbf{139}, B963 (1965). 
	
	\bibitem{Fleming2} 
	G. N. Fleming, Philos. Sci. \textbf{67}, 515 (2000).

	\bibitem{Halvorsen2} 
	H. Halvorsen, Philos. Sci. \textbf{68}, 111 (2001).
	
	\bibitem{Shojai} 
	A. Shojai and M. Golshani, Ann. Fond. Louis de Broglie {\bf 22}, 373 (1997).
	
	\bibitem{Wightman}
	A. S. Wightmann, Rev. Mod. Phys. {\bf 34}, 845 (1962).
	
	\bibitem{JandP} 
	J. M. Jauch and C. Piron, Helv. Phys. Acta \textbf{40}, 559 (1967).

	\bibitem{Amrein} 
	W. O. Amrein, Helv. Phys. Acta \textbf{42}, 149 (1969).
	
	\bibitem{Hawton1} 
	M. Hawton, Phys. Rev. A \textbf{59}, 954 (1999). 
	
	\bibitem{Hawton2}
	M. Hawton and W. E. Baylis, Phys. Rev. A \textbf{64}, 012101 (2001).
		
	\bibitem{BB2}
	I. Bia\l ynicki-Birula, Acta Phys. Pol. A {\bf 86}, 97 (1994).
	
	\bibitem{BB3}
	I. Bia\l ynicki-Birula, Progress in Optics XXXVI, pp.~245-294, Elsevier (Amsterdam 1996).

	\bibitem{Sipe}
	J. E. Sipe, Phys. Rev. A {\bf 52}, 1875 (1995).

	\bibitem{Raymer}
	B. J. Smith and M. G. Raymer, New J. Phys. {\bf 9}, 414 (2007).
		
	\bibitem{Glauber1}
	R. J. Glauber, Phys. Rev. {\bf 130}, 6 (1963).
		
	\bibitem{Knight} 
	J. M. Knight, J. Math. Phys. \textbf{2}, 459 (1961).
	
	\bibitem{Licht} 
	A. L. Licht, J. Math. Phys. \textbf{4}, 1443 (1963).
		
	\bibitem{BB1}
	I. Bia\l ynicki-Birula and Z. Bia\l ynicka-Birula, Phys. Rev. A \textbf{79}, 032112 (2009).

	\bibitem{Gross} 
	L. Gross, J. Math. Phys. \textbf{5}, 687 (1964).
		
	\bibitem{Hawton3}
	M. Hawton, Phys. Rev. A \textbf{75}, 062107 (2007).
	
	\bibitem{Hawton4} 
	M. Hawton, Proc. SPIE {\bf 6664}, 666408 (2007).
	
	\bibitem{Brody}
	D. C. Brody, J. Phys. A \textbf{47}, 3 (2013).
	
	\bibitem{Hawton5} 
	M. Hawton and V. Debierre, Phys. Lett. A. {\bf 381}, 1926 (2017).

	\bibitem{Jaromir}
	M. Dobrski, M. Przanowski, J. Tosiek and F. J. Turrubiates, {\em Uniqueness of the photon position operator with commuting components}, arXiv:2205.04791 (2022).

	\bibitem{Mostafazadeh2}  
	A. Mostafazadeh, J. Math. Phys. \textbf{43}, 205 (2002). 
	
	\bibitem{Mostafazadeh1} 
	A. Mostafazadeh,  J. Math. Phys. {\bf 44}, 974 (2003).

	\bibitem{Landau and Peierls}
	L. D. Landau and R. Peierls, Z. Phys \textbf{62}, 188 (1930).
	
	\bibitem{Cook1} 
	R. J. Cook, Phys. Rev. A \textbf{25}, 2164 (1982).

	\bibitem{Cook2} 
	R. J. Cook, Phys. Rev. A \textbf{26}, 2754 (1982).
		
	\bibitem{Mandel} 
	L. Mandel, Phys. Rev. \textbf{144}, 1071 (1966).
		
	\bibitem{exp}
	M. B. Nasr, S. Carrasco, B. E. A. Saleh, A. V. Sergienko, M. C. Teich, J. P. Torres, L. Torner, D. S. Hum, and M. M. Fejer, Phys. Rev. Lett. {\bf 100}, 183601 (2008).
	
	\bibitem{exp2}
	A. Tanaka, R. Okamoto, H. H. Lim, S. Subashchandran, M. Okano, L. Zhang, L. Kang, J. Chen, P. Wu, T. Hirohata, S. Kurimura, and S.Takeuchi , Opt. Express {\bf 20}, 25228 (2012). 
	
	\bibitem{exp3}
	M. Okano, H. H. Lim, R. Okamoto, N. Nishizawa, S. Kurimura, and S. Takeuchi, Sci. Rep. {\bf 5}, 18042 (2015). 
	
	\bibitem{exp4}
	U. A. Javid, J. Ling, J. Staffa, M. Li, Y. He and Q. Lin, Phys. Rev. Lett. {\bf 127}, 183601 (2021). 
		
	\bibitem{Bennett}
	R. Bennett, T. M. Barlow, and A. Beige, Eur. J. Phys. {\bf 37}, 014001 (2016).

	\bibitem{Szameit}
	M. Ornigotti, C. Conti and A Szameit, J. Opt. {\bf 20}, 065201 (2018).
	
	\bibitem{book}
	H. E. Hernandez-Figueroa, M. Zamboni-Rached and E. Recami, {\em Localized Waves}, John Wiley \& Sons, Inc. (Hoboken, 2008).

	\bibitem{Aiello}
	A. Aiello, J. Opt. {\bf 22}, 014001 (2020).
		
	\bibitem{Aiello2}		
	A. Aiello, J. Opt. {\bf 22}, 014002 (2020).
								
	\bibitem{Halvorsen1} 
	H. Halvorsen and R. Clifton, Philos. Sci. \textbf{69}, 1 (2002).

	\bibitem{Hegerfeldt2} 
	G. C. Hegerfeldt, Phys. Rev. D \textbf{10}, 3320 (1974).
	
	\bibitem{Hegerfeldt3} 
	G. C. Hegerfeldt, Nucl. Phys. B Proc. Suppl. \textbf{6}, 231 (1989).

	\bibitem{Skagerstam} 
	B. S. K. Skagerstam, Int. J. Theor. Phys. \textbf{15}, 213 (1976).
	
	\bibitem{Perez} 
	J. Fernando Perez and I. F. Wilde, Phys. Rev. D \textbf{16}, 315 (1977).

	\bibitem{Malament} 
	D. B. Malament, {\em In Defense of Dogma: Why There Cannot be a Relativistic Quantum Mechanics of (Localizable) Particles}. In: R. Clifton (eds) {\em Perspectives on Quantum Reality}, The University of Western Ontario Series in Philosophy of Science (A Series of Books in Philosophy of Science, Methodology, Epistemology, Logic, History of Science, and Related Fields), vol {\bf 57}  Springer Verlag, (Dordrecht, 1996).

	\bibitem{Hegerfeldt1} 
	G. C. Hegerfeldt, Phys. Rev. Lett. \textbf{72}, 596 (1994).
		
	\bibitem{Hegerfeldt4} 
	G. C. Hegerfeldt, Ann. Phys. \textbf{7}, 716 (1998).

	\bibitem{Fermi} 
	E. Fermi, Rev. Mod. Phys. \textbf{4}, 87 (1932).

	\bibitem{Shirokov} 
	M. I. Shirokov, Yad. Fiz. \textbf{4}, 1077 (1966).
	
	\bibitem{Rubin}  
	M. H. Rubin, Phys. Rev. D \textbf{35}, 3836 (1987).
		
	\bibitem{Biswas} 
	A. K. Biswas, G. Compagno, G. M. Palma, R. Passante and F. Persico, Phys. Rev. A \textbf{44}, 798(E) (1991).
	
	\bibitem{Milonni1} 
	P. W. Milonni, D. F. V. James and H. Fearn, Phys. Rev. A \textbf{52}, 1525 (1995).
	
	\bibitem{Borrelli} 
	M. Borrelli, C. Sab\'in, G. Adesso, F. Plastina and S. Maniscalco, New J. Phys. \textbf{14}, 103010 (2012).

	\bibitem{Yngvason}
	D.~Buchholz and J.~Yngvarson, Phys. Rev. Lett. \textbf{73}, 613 (1994).

	\bibitem{Ben-Benjamin} 
	J. S. Ben-Benjamin and L. Cohen, Phys. Lett. A \textbf{384}, 18 (2020).
	
        \bibitem{Dirac}
        P. A. M. Dirac, {\em The Principles of Quantum Mechanics}, Oxford University Press (Oxford, 1958).

	\bibitem{Jake} 
	J. Southall, D. Hodgson, R. Purdy, and A. Beige, J. Mod. Opt. {\bf 68}, 647 (2021).

	\bibitem{Barcellona} 
	P. Barcellona, R. Bennett and S. Y. Buhmann, J. Phys. Commun. \textbf{2}, 035027 (2018).
	
	\bibitem{Howell} 
	K. B. Howell, \textit{Principles of Fourier Analysis},  pp. 278-279 (Chapman and Hall/CRC 2001).
		
	\bibitem{Heitler}
	W. Heitler, {\em The Quantum Theory of Radiation}, Dover Publications (New York 1953).

	\bibitem{Weinberg}
	S. Weinberg, {\em The Quantum Theory of Fields}, Vol. {\bf 1}, Cambridge University Press (Cambridge, 1995).

	\bibitem{Griffiths}
	D. J. Griffiths, {\em Introduction to electrodynamics} (fourth edition), Cambridge University Press (Cambridge, 2017). 

	\bibitem{Hawton6} 
	M. Hawton, Phys. Rev. A \textbf{100}, 012122 (2019).  
	
       \bibitem{Jake2}
       J. Southall, D. Hodgson, R. Purdy and A. Beige, {\em Comparing Hermitian and non-Hermitian Quantum Electrodynamics}, arXiv:2208.01532 (2022).

	\bibitem{Allcock}
	G. R. Allcock, Ann. Phys. (N. Y.) {\bf 53}, 253 (1969).

	\bibitem{Allcock2}
	G. R. Allcock, Ann. Phys. (N. Y.) {\bf 53}, 286 (1969).
	
	\bibitem{Allcock3}
	G. R. Allcock, Ann. Phys. (N. Y.) {\bf 53}, 311 (1969).

	\bibitem{Fabio}
         M. Conforti, A. Marini, T. X. Tran, D. Faccio and F. Biancalana, Opt. Express {\bf 21}, 31239 (2013).

	\bibitem{Dickinson}
	R. Dickinson, J. Forshaw and P. Millington, Phys. Rev. D {\bf 93}, 065054 (2016).

	\bibitem{Bostelmann}
	H. Bostelmann and D. Cadamuro, Phys. Rev. D {\bf 93}, 065001 (2016).

	\bibitem{newHawton}
	M. Hawton, {\em Photon quantum mechanics in real Hilbert space},  arXiv:2107.03262 (2021).

	\bibitem{Maybee}
	B. Maybee, D. Hodgson, A. Beige and R. Purdy, Entropy {\bf 21}, 844 (2019). 
\end{thebibliography}
\end{document}